\documentclass{aa}

\def\puncspace{\ifmmode\,\else{\ifcat.\C{\if.\C\else\if,\C\else\if?\C\else%
\if:\C\else\if;\C\else\if-\C\else\if)\C\else\if/\C\else\if]\C\else\if'\C%
\else\space\fi\fi\fi\fi\fi\fi\fi\fi\fi\fi}%
\else\if\empty\C\else\if\space\C\else\space\fi\fi\fi}\fi}
\def\SP{\let\\=\empty\futurelet\C\puncspace}
\def\0{{$\clubsuit$}}
\newbox\grsign \setbox\grsign=\hbox{$>$} \newdimen\grdimen
\grdimen=\ht\grsign
\newbox\simlessbox \newbox\simgreatbox
\setbox\simgreatbox=\hbox{\raise.5ex\hbox{$>$}\llap
     {\lower.5ex\hbox{$\sim$}}}\ht1=\grdimen\dp1=0pt
\setbox\simlessbox=\hbox{\raise.5ex\hbox{$<$}\llap
     {\lower.5ex\hbox{$\sim$}}}\ht2=\grdimen\dp2=0pt
\def\simgreat{\mathrel{\copy\simgreatbox}}
\def\simless{\mathrel{\copy\simlessbox}}
\newbox\simppropto
\setbox\simppropto=\hbox{\raise.5ex\hbox{$\sim$}\llap
     {\lower.5ex\hbox{$\propto$}}}\ht2=\grdimen\dp2=0pt

\usepackage{graphics}

\begin{document}

   \thesaurus{03 (03.13.6; 11.01.1; 11.06.2; 11.09.4; 11.19.2)} 
   \title{The Emission Line Sequence of Normal Spiral Galaxies}

%   \subtitle{}

   \author{L. Sodr\'e Jr.
          \inst{1}
          \and
           G. Stasi\'nska
           \inst{2}
          }

   \offprints{L. Sodr\'e Jr.}

   \institute{Departamento de Astronomia, Instituto Astron\^omico 
              e Geof\'\i sico da USP, Av. Miguel Stefano 4200, 04301-904 
              S\~ao Paulo, Brazil \\
              email: laerte@iagusp.usp.br
         \and
              DAEC, Observatoire de Meudon, F-92195 Meudon Principal
Cedex,
              France \\
             email: grazyna@obspm.fr
             }

   \date{Received ; accepted }

\titlerunning{Emission Lines in Normal Spirals}
   \maketitle

   \begin{abstract}

We have analyzed the emission line properties in the integrated 
spectra of 15 normal spiral galaxies.  We show that very clear trends 
appear when plotting relevant emission line ratios or equivalent 
widths as a function of galaxy spectral types, obtained with a
Principal Component Analysis of the continua and absorption features
of spectra.  The equivalent widths 
of all the lines analyzed correlate extremely well with spectral types, 
implying that each of them can be considered a good indicator of the 
spectral type in normal galaxies.  The position of most galaxies of
our sample in classical emission line diagnostic diagrams follows that 
of individual giant HII regions in spiral galaxies, but for the 
earliest type galaxies, the emission line pattern resembles more that 
of LINERs.  Therefore, the direct interpretation of equivalent widths 
in terms of star formation rates would be misleading in such cases. 
The observed trends in the emission line ratios as a 
function of galaxy spectral type suggest a decrease of O/H, a decrease 
of N/O, an increase of the average effective temperature or ionization 
parameter, and a decrease of the  effective internal extinction of
galaxies  with increasing (early to late) spectral type.

   \end{abstract}

%
%________________________________________________________________

\section{Introduction}
Emission lines in normal galaxies are powerful tracers of the physical 
processes
associated with the galaxy interstellar medium and the formation of
massive
stars. They allow  quantitative estimates of the star formation rates,
 as well as of the chemical abundances and physical conditions 
of the gas in galaxies  (McCall et al.  \cite{mcc}; 
Gallagher et al.  \cite{gall}; Kennicutt, Tamblyn \& 
Congdon \cite{k94}; 
Zaritsky et al. \cite{z1}; Barbaro \& Poggianti \cite{bar}). 

A census  of emission line properties in integrated spectra  
of normal galaxies is  particularly 
interesting in view of studies of galaxies at large redshifts, too distant 
to be spatially resolved. It provides useful information also for 
nearby galaxies, as it allows investigating the global 
properties of their emission. It is well known that emission lines tend 
to become more prominent as one goes from early to late galaxy types, and 
the degree to which the spiral arms are resolved into individual HII
regions
(where most of the emission lines in normal galaxies are produced)
is one of the most important criteria for discriminating
among the spiral types in the Hubble system (Hubble \cite{hu}; Sandage 
\cite{san}). 
Kennicutt (1992a, hereafter \cite{k92a}) has already 
discussed some properties of the emission lines in integrated spectra,
mainly
evaluating their reliability as quantitative tracers of the total massive 
star formation rate in galaxies. 
Lehnert  \&  Heckman (\cite{leh}) have examined 
the location of integrated spectra of galaxies in the standard 
emission line
diagnostic diagrams used to classify emission line objects (Baldwin,
Phillips \& Terlevich \cite{bal}; Veilleux \&  Osterbrock \cite{ve}), 
concluding that the integrated emission in star-forming galaxies
has a substantial
contribution of a diffuse component, with physical properties different
from those found in high surface brightness HII regions. 
Most studies, however, have focused on the
properties of emission lines produced either in HII regions in the 
spiral arms of disk galaxies 
(e.g. McCall et al. \cite{mcc}; Belley \& Roy \cite{bel}; 
Oey \& Kennicutt \cite{oe}; Zaritsky et al. \cite{z1};
Kennicutt \& Garnett \cite{k96}; Roy \& Walsh \cite{roy}) 
or in galactic nuclei  
(Heckman et al. \cite{he}; Keel \cite{keel}; 
Ho et al. \cite{ho}). 
In particular, Zaritsky et al. (\cite{z1}) have investigated 
the oxygen abundance properties of a sample of disk galaxies from 
the spectra of individual HII regions located at various 
galactocentric radii. They found that the characteristic abundance of 
the galaxies correlates well with both their
morphological type and their luminosity.

In this paper we examine the trends of emission line 
properties in the {\em integrated} spectra of 15 normal, nearby spiral 
galaxies.
The spectra come from Kennicutt's (1992b, hereafter \cite{k92b}) 
spectro-photometric atlas. 

The galaxies are ordered by galaxy spectral type (hereafter ST), 
obtained using Principal Component Analysis of the integrated properties 
of the continua and absorption features of 23 normal galaxies of all
morphologies, in a manner similar to the one 
introduced by Sodr\'e \& Cuevas (\cite{s94},
\cite{s97}) and several other authors
(e.g.,
Connolly et al. \cite{co}; Zaritsky et al. 
\cite{z2}; Folkes et al. \cite{fo}; 
Galaz \& de Lapparent \cite{gala}). 
Such a procedure allows to
define a spectral classification that correlates well with Hubble
morphological types and present some advantages 
over the usual morphological classification. Firstly, it provides
quantitative, continuous,
and well defined types, avoiding the ambiguities of the 
intrinsically more qualitative and subjective morphological
classification.
Secondly, the sequence of galaxies obtained in this way is
easier to model than the Hubble sequence, 
because the physical process behind galaxy spectra are better
understood than those needed to explain galaxy morphologies. Thirdly, 
this classification can be
applied to redshift surveys where no information is available on the 
galaxy morphologies, allowing to
``recycle" data obtained for other purposes (de Lapparent, Galaz \&
Arnouts
\cite{lap}; Bromley et al. \cite{bro}).

We will show that the  patterns of variation of emission line properties 
along the sequence of normal galaxies are much more regular when 
considering the galaxy spectral 
type sequence than when using the morphological sequence.

This paper is organized as follows. In Sect. 2 we introduce the  
sample of normal galaxies, briefly describe the 
procedure used to obtain spectral types from integrated galaxy spectra,
and
 present the relevant data for the 
 emission lines in the 
blue/visible region of the spectra.
In Sect. 3 we present the trends of equivalent widths and emission
line ratios with galaxy spectral types.
In Sect. 4 we interpret these trends, after a brief reminder on 
emission line theory and a comparison of the loci of integrated galaxy 
spectra with those of giant HII regions in classical diagnostic 
diagrams. Finally, in Sect. 5 we summarize our results.

\section{The data base}

\subsection{The sample of galaxies}

The integrated spectra of galaxies discussed in this work come from 
the atlas of Kennicutt 
(\cite{k92b}). They are suitable for setting the basis of a global
characterization of galaxies because this is the largest sample of
integrated spectra of low-redshift galaxies available and obtained in 
a uniform way. From this atlas we, conservatively, 
rejected all the galaxies showing any evidence of strong spectral
peculiarities or interactions (i.e. AGNs, starbursts and 
mergers). This left us with a total of 23 normal galaxies, which 
 will be our standards. They are presented 
in Table 1, together with their Hubble types (taken
from  \cite{k92a}), their T-types (following the RC3 convention-
de Vaucouleurs et al. \cite{vauc}), their spectral types (see below), and
the 
spectral classification of their nuclei according to Ho et al. 
(\cite{ho}; see Sect. 4). These galaxies are all at high Galactic
latitude, so foreground absorption by our Galaxy is unimportant.
Note that the set of standards adopted here is very similar to those
used to characterize normal galaxies in other studies of spectral 
classification (e.g. Folkes et al. \cite{fo}; Galaz \& de Lapparent 
\cite{gala}).

\begin{table*}
\caption{Standard galaxies and classification}
\begin{tabular}{lcccc}
\hline
{ Name}&{Hubble type}&{$T$-type}&{spectral type} & {nuclear types} \\
NGC\,3379  & E0    &  -5  & -5.2  & \\
NGC\,4472  & E1/S0 &  -4  & -5.7  & \\ 
NGC\,4648  & E3    &  -5  & -4.4  & \\ 
NGC\,4889  & E4    &  -5  & -4.1  & \\ 
NGC\,3245  & S0    &  -2  & -4.6  & \\ 
NGC\,3941  & SB0/a &  ~0  & -2.7  & \\ 
NGC\,4262  & SB0   &  -2  & -3.8  & \\ 
NGC\,5866  & S0    &  -2  & -4.4  & \\ 
NGC\,1357  & Sa    &  ~1  & -2.0  & \\ 
NGC\,2775  & Sa    &  ~1  & -3.4  & \\
NGC\,3368  & Sab   &  ~2  & -3.3  & L2 \\ 
NGC\,3623  & Sa    &  ~1  & -4.1  & L2: \\
NGC\,1832  & SBb   &  ~3  & ~2.4  & \\
NGC\,3147  & Sb    &  ~3  & ~0.0  & S2 \\
NGC\,3627  & Sb    &  ~3  & ~1.4  & T2/S2 \\ 
NGC\,4775  & Sc    &  ~5  & ~9.7  & \\
NGC\,5248  & Sbc   &  ~4  & ~1.4  & H \\ 
NGC\,6217  & SBbc  &  ~4  & ~4.8  & H \\ 
NGC\,2903  & Sc    &  ~5  & ~1.1  & H \\ 
NGC\,4631  & Sc    &  ~5  & ~9.4  & H \\
NGC\,6181  & Sc    &  ~5  & ~3.1  & H \\
NGC\,6643  & Sc    &  ~5  & ~3.8  & H \\
NGC\,4449  & Sm/Im &  ~9  & 10.8  & H \\
\hline
\end{tabular}
\end{table*}

\subsection{The spectral classification}
 
Our general approach to spectral classification is described in detail in 
Sodr\'e \& Cuevas (\cite{s97}) and in Cuevas et al. (\cite{cu}). Here
we only present an overview of the procedures used to obtain  spectral
types for the galaxies discussed here. The point to be stressed is that
the spectra of normal galaxies form a sequence- the spectral sequence- in
the spectral space spanned by the $M$-dimensional vectors that contain the
spectra, each vector being the flux of a galaxy (or a scaled version of
it)
sampled at $M$ 
wavelengths (Sodr\'e \& Cuevas \cite{s94}, \cite{s97}; 
Connolly et al. \cite{co}; Folkes et al. 
\cite{fo}). The spectral sequence correlates well with the Hubble
morphological sequence, and we define the spectral type of a galaxy from
its
position along the spectral sequence.

The spectral sequence is identified using Principal Component Analysis
(PCA).
PCA allows to define a new orthonormal reference system in spectral 
space, centered at the baricenter of galaxy spectra and with 
basis-vectors (the principal components) spanning directions of maximum
variance. The plane defined by the first two principal components- which
we
call principal plane-   is the plane that 
contains the maximum variance in the spectral space.

PCA was applied to a pre-processed version of the 23 original galaxy
spectra.  
Firstly, the spectra were shifted to the rest frame and re-sampled in 
the wavelength interval from 3784 \AA~ to 6500 \AA, in equal-width bins of
2 \AA.
Secondly, we removed from the analysis 4 regions
of $\sim$20 \AA~each centered at the wavelengths of He II $\lambda$4686,
H$\beta$ $\lambda$4861, [O III] $\lambda$4959 and [O III] $\lambda$5007. 
This was done in order to avoid the inclusion of emission lines in the 
analysis, that increases the dispersion of the spectra in the principal 
plane (mainly due to an increase in the second principal component).
The spectra, now sampled at $M=1277$ wavelength intervals, were then 
normalized to the same mean flux ($\sum_\lambda f_\lambda= 1$). 
After, we computed a mean spectrum and, finally, we subtracted this mean
spectrum from the spectrum of each galaxy. 
PCA  was then used to obtain 
the principal components.  
This procedure is equivalent to the
conventional PCA on the covariance matrix, that is, the basis vectors are
the eigenvectors of a covariance matrix. 
Note that the principal components are dependent of the data and its
pre-processing, and using a different normalization for the spectra or
taking some other wavelength interval  certainly would lead to a different 
result for the values of the principal components. However, it
can be shown that, to a large extent, these solutions are equivalent, 
all producing essentially the same ranking of galaxies along the first 
principal component. 

Fig. 1 shows the projection of the spectra of the 23 standards on to the
principal plane. This plane is highly informative about the structure of
the spectral space, as it contains 94.8\% of the data variance.
The spectral  sequence is readily identified along the first principal
component. We then define quantitatively the spectral type ST of a galaxy 
by the value of its first principal component. These values are presented 
in table 1.  Note that ST increases from early to late types.
In what follows, the
spectral sequence will refer to that obtained with the 23 standards of
table 1.
The mean values and dispersions of ST for the standards
are -4.9 $\pm$ 0.8 for E-E/S0, -3.9 $\pm$ 0.9  for S0-S0/a, -3.2 $\pm$ 0.9  
for Sa-Sab, 2.0 $\pm$ 1.8  for Sb-Sbc, and 6.3 $\pm$ 4.1  for Sc-Im.
The ranking of spectral types in these groups is analogous to the 
ranking of Hubble morphological types. Note that the 
dispersion of ST around the mean value of each group is large enough to 
produce a substantial overlap in the spectral sequence of galaxies of 
different Hubble types: galaxies of same 
morphologies may have very different spectral types.

\begin{figure}
%\resizebox{\textwidth}{!}{\includegraphics{figure1.ps}}
%\begin{figure}
\resizebox{\columnwidth}{!}{\includegraphics{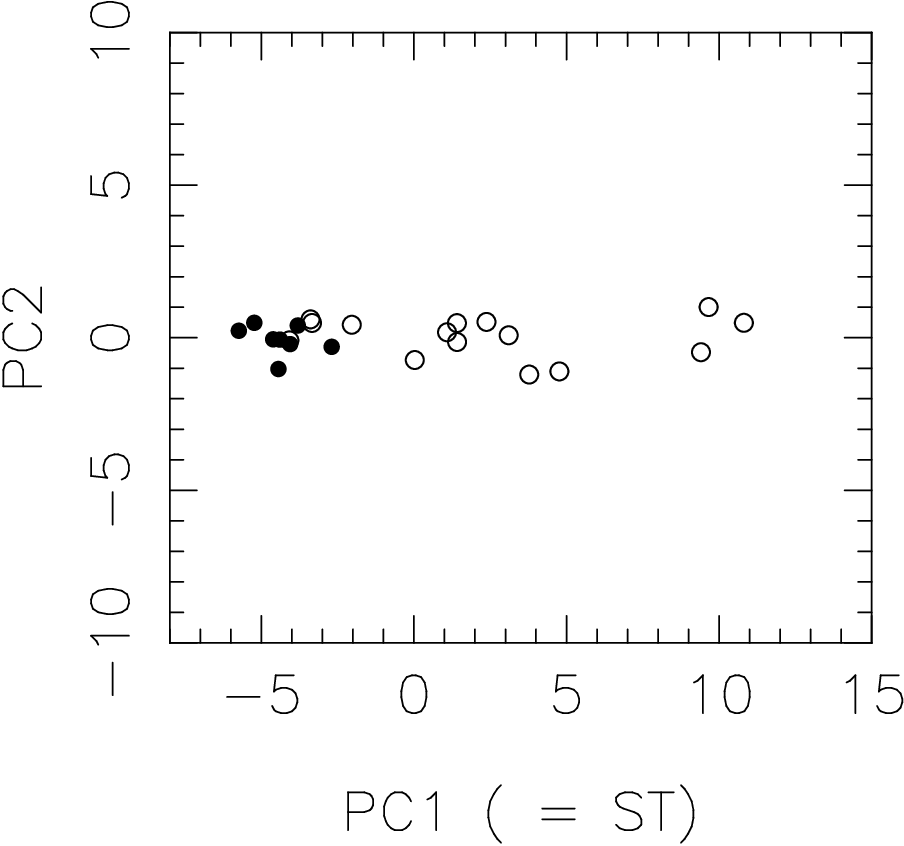}}
      \caption[]{
Projection of the spectra of the 23 standard galaxies
on to the principal plane. The spectral sequence follows the first
principal
component, and a spectral type is attributed to each galaxy by its
value of this component. Filled symbols: E and S0 galaxies; open symbols:
S galaxies.             }
         \label{f1}
\end{figure}

It is worth emphasizing that we
base our spectral classification only on the integrated properties of 
the stellar populations that are contained in the continuum and
absorption lines, and that the emission lines enter in no way in the 
classification scheme.

\subsection{Emission line data and reddening correction}

Among our 23 standard galaxies, emission lines are measured only 
for the 15 spirals ($T \ge 1$), being too weak or undetected in the
integrated spectra of galaxies of earlier types (at the resolution of
the observations, $\sim$5-7\AA; see \cite{k92b}). 
We use all the emission lines with equivalent widths (EW) measured by
\cite{k92a}:
[OII]$\lambda$3727, H$\beta$, [OIII]$\lambda$5007, H$\alpha$,
[NII]$\lambda 
\lambda$6548, 6583, and [SII]$\lambda \lambda$6717, 6731. The EW of
H$\alpha$
was computed from the EW of H$\alpha$+[NII] and the value of
[NII]/H$\alpha$,
also from \cite{k92a}. Other emission lines,
like [OI]$\lambda\lambda$6300,6363, are not strong enough to be reliably
measured in the spectra of Kennicutt's atlas (\cite{k92b}).

In the next sections we will discuss the behavior of reddening 
corrected line ratios.  For that, we
have obtained a value of the effective internal reddening of the
galaxies from an estimation of the emission line
intensity ratio of H$\alpha$ and H$\beta$,
H$\alpha$/H$\beta$. Our procedure is similar to the one adopted by \cite{k92a}.
We begin by modeling the relation between the observed 
EWs of H$\alpha$ and H$\beta$ by a straight line (Fig. 2).
Using the bisector estimation for the fitting
(Isobe et al. \cite{is}), we have obtained
\begin{eqnarray}
 {\rm EW(H}\beta)_{obs} =&-4.19 (\pm 0.36)+ \nonumber \\
& 0.194 (\pm 0.011) {\rm EW(H}\alpha)_{obs}
\end{eqnarray}
Note that this linear relationship is purely empirical, and there is not,
{\it a priori}, any reason for EW(H$\beta)_{obs}$ and 
EW (H$\alpha)_{obs}$ be linearly related. 

The observed EW of a line may be 
written as EW$_{obs}$ = EW$_{em}$ + EW$_{abs}$, where 
EW$_{em}$ is the emission 
component ($> 0$) and EW$_{abs}$ corresponds
to the component due to stellar absorption ($< 0$).
To proceed, we need to know how EW$_{abs}$ behaves as a function of EW$_{em}$ 
or EW$_{obs}$. 
Let us suppose, then, that the emission and observed EWs of
H$\alpha$ and H$\beta$ may also be related linearly,  
what is a reasonable assumption in most models 
(e.g., Barbaro \& Poggianti \cite{bar}). 
Writing
EW(H$\alpha)_{em} = a_\alpha + b_\alpha $EW(H$\alpha)_{obs}$ and
EW(H$\beta)_{em} = a_\beta + b_\beta $EW(H$\beta)_{obs}$, and substituting
these relations in Eq. 1, we obtain 
EW(H$\beta)_{em} = C + R ~ $EW(H$\alpha)_{em}$, where $C$ and $R$ are 
constants.
Now, physical solutions require $C$ equal to zero, because
EW(H$\beta)_{em}$ should go to zero when EW(H$\alpha)_{em}$ goes to zero.
Consequently, we obtain that the emission EWs of 
H$\alpha$ and H$\beta$ are related by
\begin{equation}
{\rm EW(H}\beta)_{em} = R ~ {\rm EW(H}\alpha)_{em} 
\end{equation}
with $R = 0.194 b_\beta/ b_\alpha$. 
With these assumptions, H$\alpha$/H$\beta$
depends only of $R$ and the ratio of the continuum fluxes at H$\alpha$ and
H$\beta$, that we have measured in the spectra. We have adopted here
$R=0.2$, since models
like those of Barbaro \& Poggianti (\cite{bar})
indicate that $ b_\beta \sim b_\alpha$. 

The extinction coefficient at H$\beta$, $C({\rm H}\beta)$, has been 
computed assuming an intrinsic value for H$\alpha$/H$\beta$
of 2.9 (corresponding to an electron temperature of 9000K in the case B 
of the theory of recombination-line radiation).
The expected formal error in $C({\rm H}\beta)$ from this procedure
is $\sim 0.18$. The values of $C({\rm H}\beta)$ may be considered a crude
estimation of the global extinction in the
galaxies, since, as mentioned above, foreground extinction by
 our Galaxy is negligible.

It is important to stress that, although very simple, this estimate of the
reddening is quite robust, and there
is not any evidence in the data that a more complex model is required.
As we shall see, it leads to a H$\alpha$/H$\beta$
ratio that decreases steadly with spectral type. We have verified that
this trend continues when other ways to estimate the reddening are used.
Also, the behavior of other line ratios discussed in the next sections are,
to a large extent, not strongly affected by the use of other values of $R$.

\begin{figure}
%\resizebox{\textwidth}{!}{\includegraphics{figure2.ps}}
\resizebox{\columnwidth}{!}{\includegraphics{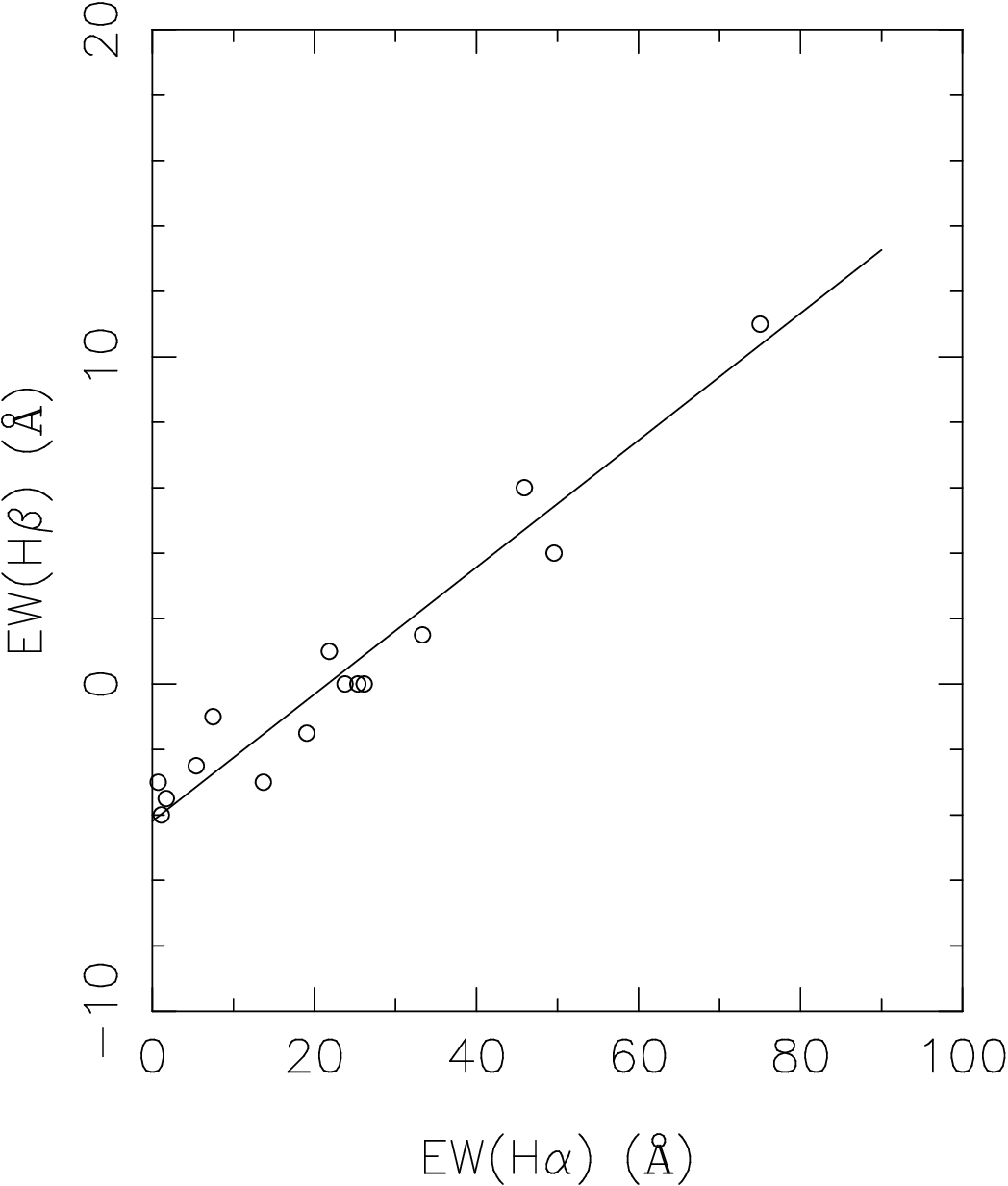}}
      \caption[]{
Relation between the observed values of the equivalent widths of
H$\alpha$ and H$\beta$. The straight line corresponds to
the fitting given by Eq. 1.}
         \label{f2}
\end{figure}

\begin{table*}
\caption{Extinction and corrected line intensities of
spiral galaxies }
\begin{tabular}{lcccccc}
\hline
Name$^{\mathrm{a}}$ & $C({\rm H}\beta)$ & [OII]/H$\beta$ &
[OIII]/H$\beta$ & [NII]/H$\alpha$ & [SII]/H$\alpha$ & 
H$\alpha$/H$\beta$  \\
NGC1357:  & ~0.71 & ~1.70 & $<$0.14 & ~0.23 & ~0.05 & ~5.12  \\ 
NGC2775:  & ~0.72 & ~3.10 & $<$1.11 & ~0.28 & - & ~5.20  \\ 
NGC3368:  & ~0.73 & ~4.76 & $<$0.74 & ~0.47 & ~0.16 & ~5.22  \\ 
NGC3623:  & ~0.72 & ~3.93 & $<$1.79 & ~0.83 & ~0.27 & ~5.18  \\
NGC1832:  & ~0.52 & ~2.21 & ~0.26 & ~0.15 & ~0.07 & ~4.42  \\ 
NGC3147:  & ~0.63 & ~0.94 & $<$0.18 & ~0.16 & - & ~4.83  \\ 
NGC3627:  & ~0.58 & ~1.05 & $<$0.08 & ~0.16 & ~0.06 & ~4.65  \\ 
NGC4775   & ~0.16 & ~3.40 & ~1.25 & ~0.08 & ~0.10 & ~3.31  \\ 
NGC5248   & ~0.58 & ~1.28 & ~0.13 & ~0.14 & ~0.06 & ~4.62  \\ 
NGC6217:  & ~0.40 & ~1.49 & ~0.26 & ~0.18 & ~0.07 & ~3.99  \\ 
NGC2903:  & ~0.64 & ~1.29 & $<$0.09 & ~0.16 & ~0.05 & ~4.84  \\ 
NGC4631   & ~0.24 & ~3.44 & ~1.20 & ~0.07 & ~0.10 & ~3.52  \\ 
NGC6181   & ~0.55 & ~1.51 & ~0.30 & ~0.14 & ~0.06 & ~4.51  \\ 
NGC6643:  & ~0.49 & ~1.35 & ~0.18 & ~0.13 & ~0.07 & ~4.31  \\ 
NGC4449   & ~0.16 & ~3.52 & ~1.82 & ~0.04 & ~0.08 & ~3.31  \\ 
\hline
\end{tabular}

$^{\mathrm{a}}$ The sign `:' after the galaxy name indicates that 
EW(H$\beta)_{obs} \le 0$ 
\end{table*}

\section{Trends of emission line properties with galaxy spectral type }

\subsection{Equivalent widths}

\begin{figure*}
%\resizebox{\textwidth}{!}{\includegraphics{figure3.ps}}
\resizebox{15.cm}{!}{\includegraphics{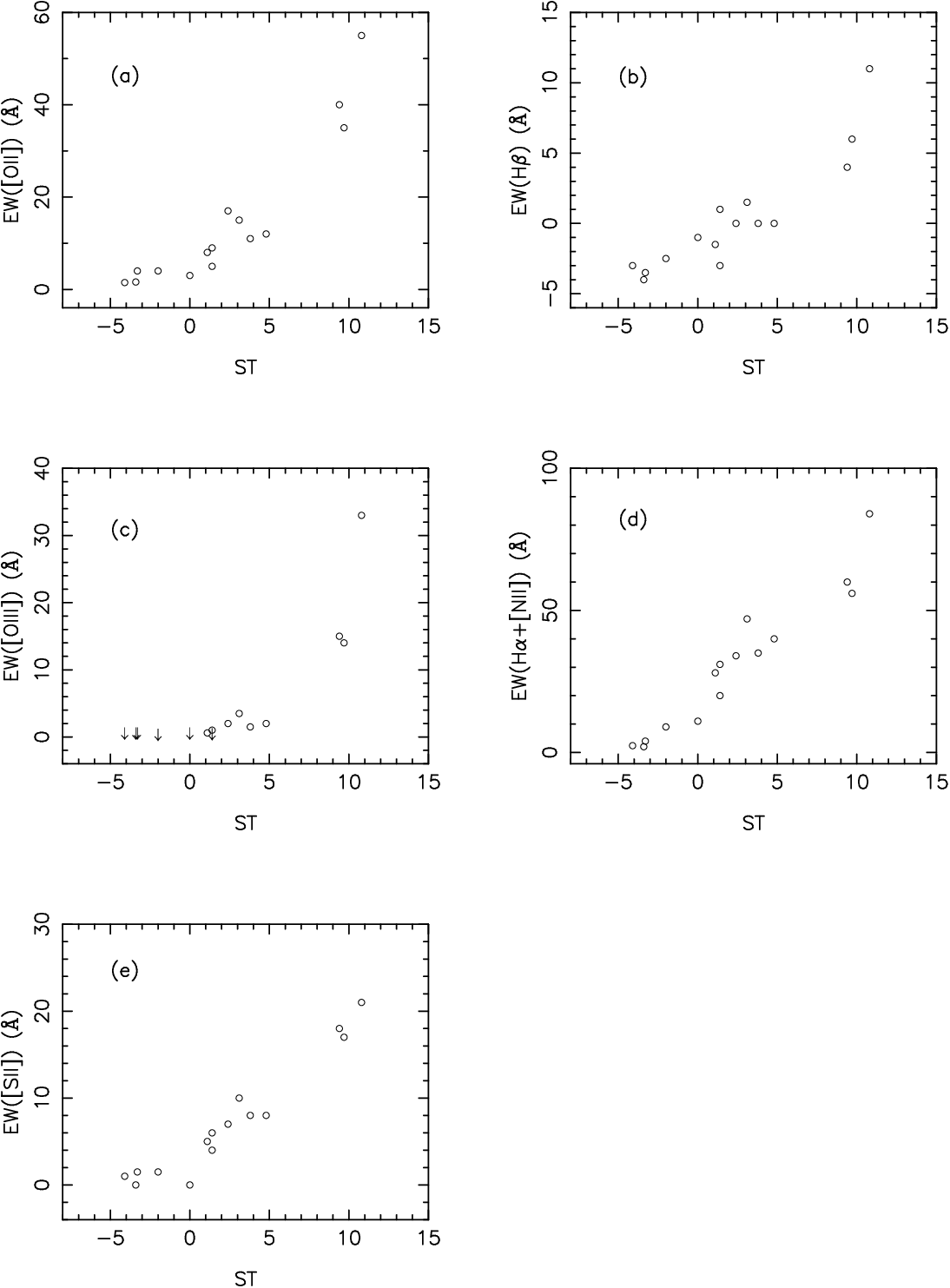}}
      \caption[]{
Relation between the observed equivalent widths of  
{\bf a}) [OII], {\bf b}) H$\beta$, {\bf c}) [OIII], {\bf d})
H$\alpha$+[NII], 
and {\bf e}) [SII] 
and galaxy spectral type (ST). Upper values of EW([OIII]) are indicated
by arrows in panel {\bf c}.
                }
         \label{f3}
\end{figure*}

 Fig. 3 consists of 5 panels showing the behavior of the observed 
 EWs of [OII], H$\beta$, [OIII], H$\alpha$+[NII] and 
 [SII] as a function of ST.  In all of them we see a clear increase of 
 the emission line EWs as a function of the galaxy 
 spectral type.  For [OIII], the increase is apparent only from ST=2 
 upwards, since this line is too weak to be effectively
 measured in spectral types 
 lower (i.e., earlier) than that.  We had shown, in Fig. 2, the 
 relation between EW(H$\beta$) and EW(H$\alpha$).  Fig. 4 shows the 
 relation of the EWs of [OII], [OIII], and [SII] with 
 EW(H$\alpha$+[NII]).  The EWs of all these lines correlate well with 
 EW(H$\alpha$) or EW(H$\alpha$+[NII]).  The EWs of the 
 Balmer emission lines are considered good tracers of the star 
 formation rate in galaxies (Kennicutt \cite{k83}), and it has been shown 
 that [OII] is a good substitute of the Balmer lines in high-redshift 
galaxies (Gallagher et al. \cite{gall}; \cite{k92a}).  Our 
 results indicate that, for practical purposes, any of the EWs
 displayed in Fig. 4 (except maybe [OIII], which is available 
 for a smaller range of galaxy types) can be used to estimate the star 
 formation rate in normal galaxies, after empirical calibration on the 
 Balmer line EWs.  It is worth noting, however, that 
 deriving a star formation rate from an emission line  equivalent 
width is model 
 dependent because, among other things, such a procedure assumes that the 
 ionization is provided by the radiation field from massive stars, 
 which is not necessarily true, specially in early type galaxies (see 
 below).  On the other side,  the EW of any of these lines is an 
 empirical indicator of galaxy spectral types, given the good 
 correlation between these two quantities, shown in Fig. 3.

\begin{figure}
%\resizebox{\columnwidth}{!}{\includegraphics{figure4.ps}}
\resizebox{8.cm}{!}{\includegraphics{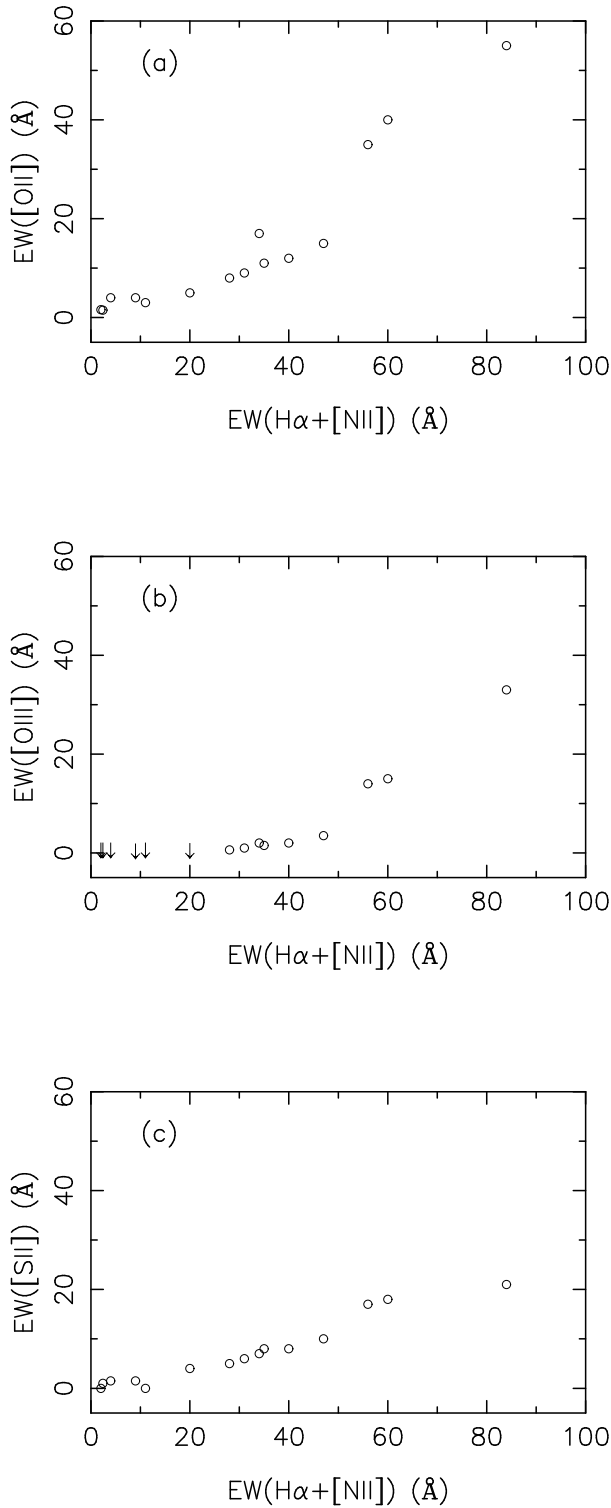}}
%\resizebox{\hsize}{!}{\includegraphics{figure4.ps}}
      \caption[]{
Relation of the observed equivalent widths of {\bf a})
[OII], {\bf b}) [OIII], and {\bf c}) [SII] with EW(H$\alpha$+[NII]).
                }
         \label{f4}
\end{figure}

It is interesting to compare our Figs. 2 and 4 with Figs. 4 to 6 of
\cite{k92a},
which include all galaxies with emission lines in \cite{k92b}.
Most of the scatter observed in the \cite{k92a} plots is produced by
galaxies not belonging to the  sample of normal galaxies discussed here.

\subsection{Emission line intensity ratios}

Fig. 5 consists of 9 panels showing the behavior of several 
emission line ratios as a function of the galaxy spectral type. 
Line ratios are corrected for reddening, as explained in Sect. 2.3.  

Fig. 5 shows that very clear trends exist between emission line 
ratios and spectral types, with very small scatter. Briefly, the 
trends can be described as follows.
\begin{enumerate}
\item{ H$\alpha$/H$\beta$ (panel 5-a)  decreases steadily with increasing 
spectral type. Converted into total extinction at H$\beta$, this means 
that C(H$\beta$) decreases from about 0.7 at ST $\sim$-5 to 0.2 at 
ST $\sim$ +10.}
\item{ [OII]/H$\beta$ (panel 5-b) first decreases with spectral type, 
for ST $\simless$ -2, then increases slowly.}
\item{ [OIII] is measurable only from ST $\sim$ 1 upwards, where 
[OIII]/H$\beta$ (panel 5-c) increases very strongly with ST. For ST
$\simless$
1 only upper values are available for the [OIII] emission but
[OIII]/H$\beta$
seems to decrease in the interval -5 $\simless$ ST $\simless$ 0.}
\item{ [NII]/H$\alpha$ (panel 5-d) decreases abruptly for ST  $<$ 0, 
decreasing mildly towards larger values of ST.}
\item{ [SII]/H$\alpha$ (panel 5-e) also decreases for ST $<$ 0, then
increases 
smoothly for larger values of ST.}
\item{ ([OII]+[OIII])/H$\beta$ (panel 5-f) first decreases until ST $\sim$
0 
(this effect is due to [OII] only since [OIII] is weak in this range of
ST), 
then increases steadily, but with a smaller slope than [OIII]/H$\beta$.
Note that, contrarily to what one might think, the values of
([OII]+[OIII])
for early-type galaxies are not upper values, since in this case the [OII]
emission is much larger than the [OIII]   emission. }
\item{ [OIII]/[OII] (panel 5-g) increases
steadily for ST $\simgreat$ 1 (the effect of [OIII] is dominant).
For lower values of ST only upper values of [OIII]/[OII] are known, but
there is a hint that this ratio decreases with ST.}
\item{ [NII]/[OII] (panel 5-h) is roughly constant  or decreases
slowly until ST $\sim$ 5, decreasing faster for larger values of ST.}
\item{ [SII]/[OII] (panel 5-i) shows little trend with respect to ST, 
except, maybe, a slight decrease  with increasing spectral types.}
\end{enumerate}

The trends seen in Figs. 3 and 5 
are impressive, considering that the galaxies 
were selected simply using a criterion of  ``normality", and 
that the ordering by spectral type relies only on the continuum and 
absorption features of the integrated spectra. 
However, if we recall that these features 
reflect the galactic stellar populations, and if the stellar populations 
obey some simple rules, then it is not so surprising that the emission
line 
properties from integrated spectra can also be ordered. Indeed, emission 
lines arise due mainly to the presence of 
hot stars ($T_{eff}$ $\simgreat$ 30000K) in the 
stellar populations, and are modulated by other parameters such as
chemical 
abundances and gas content, which also seem to be related to the general 
galaxy properties (Roberts  \&  Haynes \cite{rob}). 
Individual HII regions in a galaxy show a definite scatter in their 
properties
 (e.g., Kennicutt \& Garnett \cite{k96}, Roy \& Walsh \cite{roy}), 
but their total number in a galaxy is very large (typically of the order 
of one hundred) 
so that the integrated emission line spectrum of a galaxy is well defined. 
The fact that we find a
close link between the emission lines and the continuum and absorption
features that are behind the spectral classification suggests that 
there is a simple relation between the population of hot, ionizing 
stars and the population of stars responsible for the optical continuum. 

\begin{figure*}
\resizebox{15.cm}{!}{\includegraphics{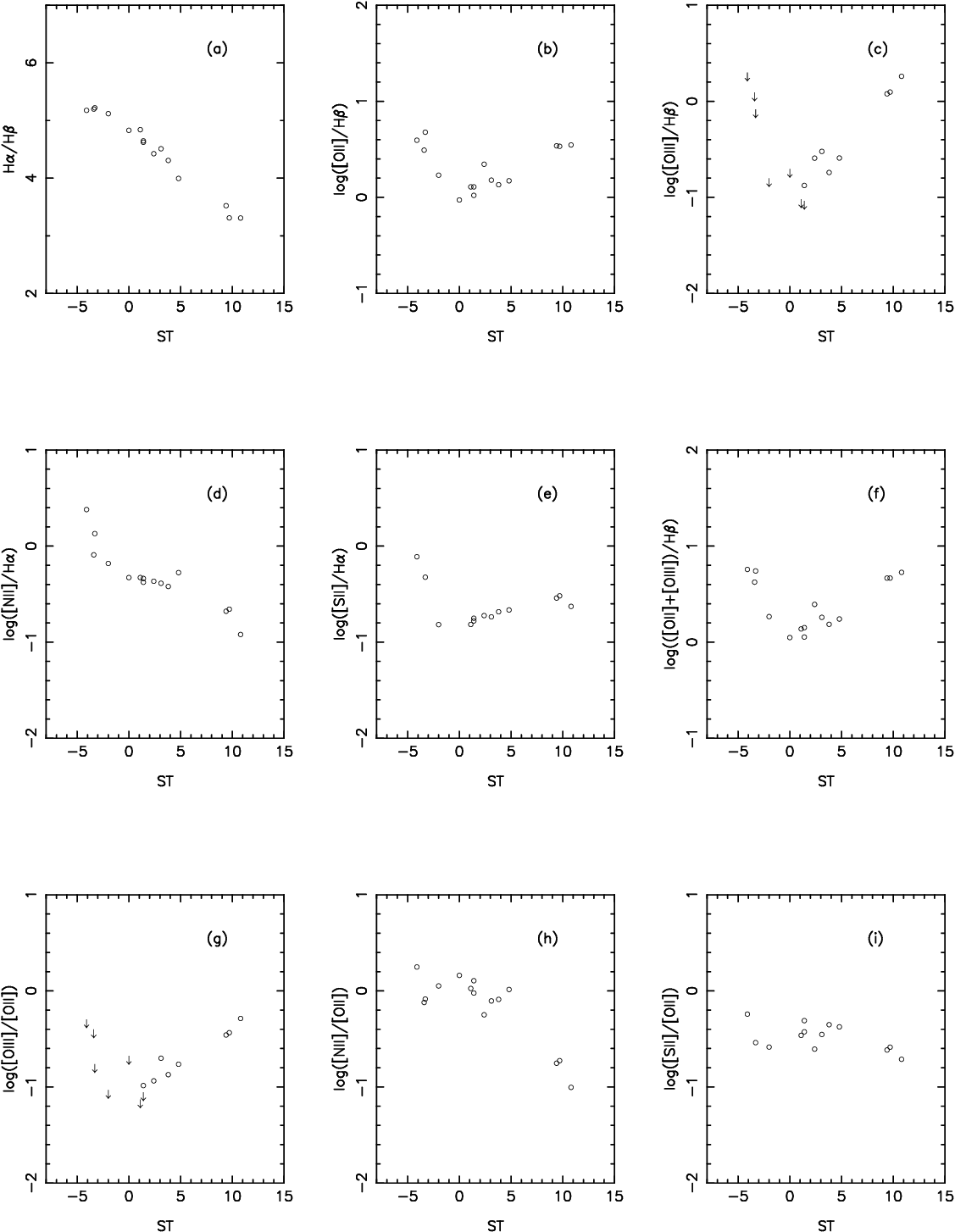}}
%\begin{figure}
%\resizebox{\columnwidth}{!}{\includegraphics{C_D_redshift.ps}}
      \caption[]{
Behavior of  emission line 
ratios measured in the galaxy integrated spectra as a function of the
galaxy 
spectral type (ST). The line ratios (except H$\alpha$/H$\beta$) 
are corrected for  reddening as described in the text. 
The emission line ratios in the Fig. are:
{\bf a})  H$\alpha$/H$\beta$, {\bf b}) [OII]/H$\beta$, 
{\bf c}) [OIII]/H$\beta$,
{\bf d}) [NII]/H$\alpha$, {\bf e}) [SII]/H$\alpha$, 
{\bf f}) ([OII]+[OIII])/H$\beta$,
{\bf g}) [OIII]/[OII], {\bf h}) [NII]/[OII], {\bf i}) [SII]/[OII].
Arrows indicate upper values.
                }
         \label{f5}
\end{figure*}

It is interesting to see how the emission line ratios behave as a function 
of the de Vaucouleurs 
type of the galaxy, which is so far the most widespread way 
of ordering galaxies and is based on  their morphological type. Fig. 6 
shows the same line ratios as Fig. 5, but as a function of the de 
Vaucouleurs $T$ type. 
One can distinguish trends similar to those seen in Fig. 5, but they 
are much less clear. 
That similar trends can be seen is not surprising, 
since there is some correlation between the spectral sequence  
and the Hubble sequence (Sodr\'e  \&  Cuevas \cite{s97}, Folkes et al.
\cite{fo}). That the trends are less clear than those 
in Fig. 5 can be explained by assuming that the populations 
of hot stars are not expected to be correlated with galaxy morphological 
types to the same extent as they are expected to be correlated with the
entire 
stellar populations, as measured by the spectral types.

\begin{figure*}
\resizebox{15.cm}{!}{\includegraphics{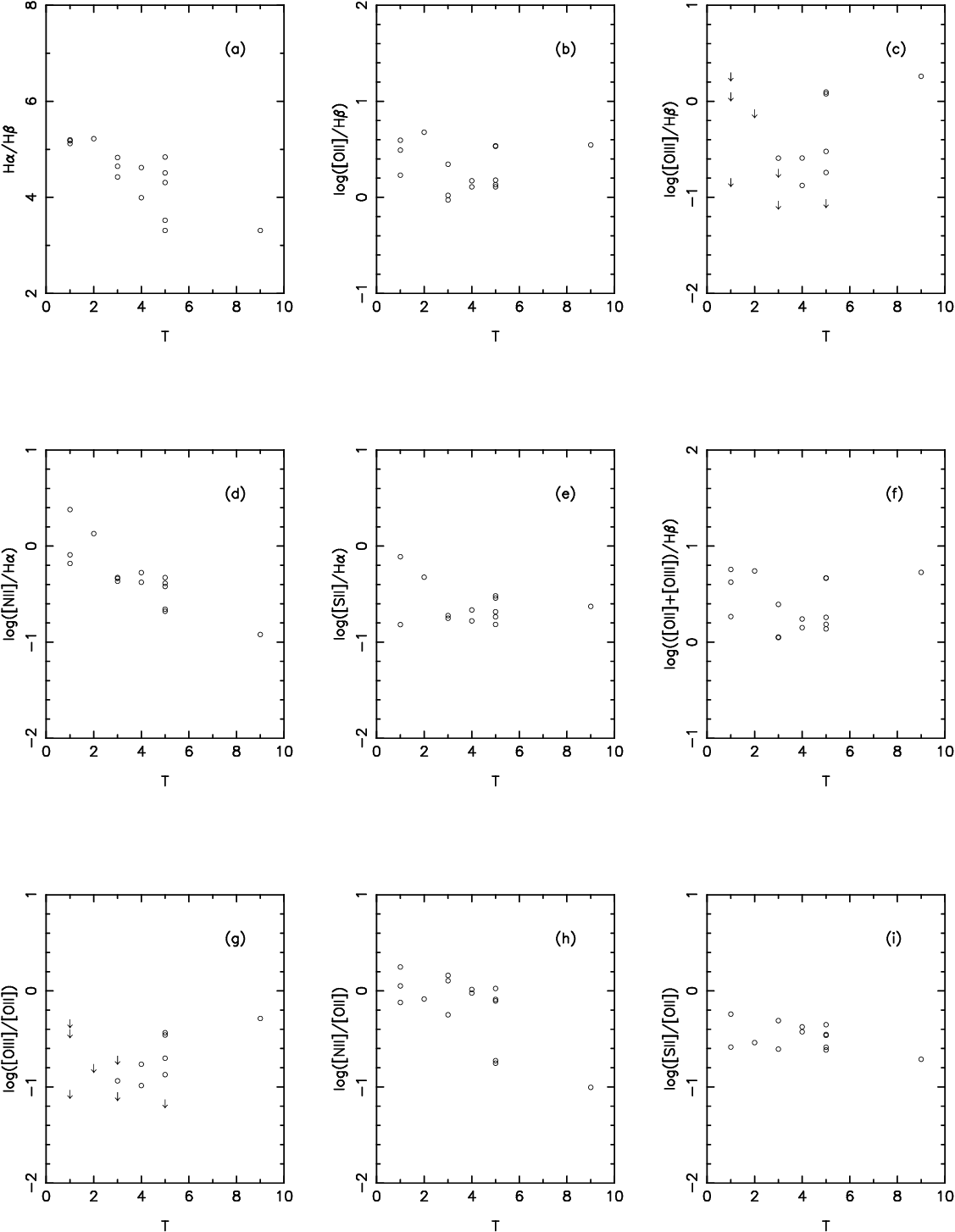}}
      \caption[]{
Same as Fig. 5, but now the line 
ratios are plotted
as a function of the de Vaucouleurs type of the galaxies. 
                }
         \label{f6}
\end{figure*}

\section{Discussion}

Before trying to interpret the trends shown above, it is useful to 
recall some basics on emission lines and photoionization models. This
is done in the next subsection.

\subsection{A reminder on emission line theory}

Schematically, the (reddening corrected)   
intensity ratios of emission lines produced in photoionized regions are 
a function of the following parameters (see e.g. Stasi\'nska \cite{st} and 
references therein):

\begin{enumerate}
\item  {the global metallicity, O/H;}
\item { abundance ratios, like N/O and S/O;}
\item {the mean effective temperature of the ionizing radiation field 
$<T_{eff}>$;}
\item  {the average ionization parameter 
$U = Q_H/(4\pi R_s^2 n)$, where $Q_H$ is the number 
of ionizing photons of an HII region, $R_s$ is the Str\"omgren radius, and 
$n$ is the gas density.}
\end{enumerate}

The effect of these parameters on the line ratios are the following.
\begin{enumerate}
\item Oxygen is the major coolant in HII regions. As O/H increases, 
the electron temperature decreases because of the increased cooling of the
gas.
The [OIII]/H$\beta$ ratio first increases (due to an abundance effect), 
then, for O/H greater than about  half  solar, decreases due to an
electron temperature effect: the gas becomes so cool that the 
optical [OIII]$\lambda$5007 line gets difficult to excite. In this case, 
cooling occurs through the [OIII] far infrared lines at 52$\mu$m and
88$\mu$m.
Qualitatively, the [OII]/H$\beta$ ratio behaves similarly to
[OIII]/H$\beta$, 
but it is not so reduced at high metallicities, because 
the cooling is less efficient in the region emitting [OII] than 
in the region emitting [OIII].
Line ratios like [NII]/H$\alpha$ or [SII]/H$\alpha$ are affected by O/H 
through the electron temperature: they get enhanced as O/H decreases.

\item  An increase of N/O only enhances [NII]/[OII], 
but does not affect the other optical line ratios,
since nitrogen contributes little to the cooling at the 
abundances expected for the general interstellar medium. 
Similarly, an increase of S/O enhances [SII]/[OII] 
and leaves the rest unchanged. 
However, one does not expected S/O to vary among galaxies,
 since sulfur and oxygen are 
produced in the same stars. Therefore, any change in [SII]/[OII] should
rather 
be attributed to other causes.

\item The effects of  $<T_{eff}>$ are twofold. 
Firstly, $<T_{eff}>$ acts on the ionization structure: 
the proportions of O$^+$, N$^+$ or S$^+$ ions  decrease with increasing  
$<T_{eff}>$.
Secondly, it influences the thermal balance of the 
nebula: as  $<T_{eff}>$ increases, the energy gains become larger and 
the electron temperature rises, increasing the intensities of the 
forbidden lines with respect to H$\alpha$ or H$\beta$. 

\item The effects of decreasing $U$ are to reduce the average ionization, 
and to decrease  ratios like O$^{++}$/O$^+$. For very low $U$, 
such as found in the diffuse interstellar medium, a wide transition 
zone of low ionization develops, containing O$^{0}$  and S$^{+}$ ions
and still hot enough to allow 
collisional excitation of optical forbidden lines. Therefore,  the 
[SII]/H$\alpha$ and the [SII]/[OII] ratios are higher than in HII regions 
having a large $U$. 
\end{enumerate}

If shocks are present, the emission line ratios are  modified. The
effect 
of shocks is to heat the gas to very high temperatures 
($T_e= 10^6 - 10^7$ K). By recombination and 
free-free processes, this produces a hard radiation field which 
strongly heats the post shock region, 
producing an extended, warm, transition region, and enhancing the lines
that 
are most sensitive to the temperature. As a 
result,  [OII]/H$\beta$, [NII]/H$\alpha$, and
[SII]/H$\alpha$ line ratios are enhanced with 
respect to pure photoionization nebulae.

\subsection{Diagnostic diagrams}

Line ratio diagrams (Baldwin et al. \cite{bal}, 
Veilleux  \& Osterbrock \cite{ve}) are helpful for the diagnostics of 
emission line objects. We now use such diagrams to compare the emission 
properties of galaxy integrated 
spectra with those of individual giant HII regions. 
Figs. 7 and 8 show the data from the integrated spectra  of the normal 
spiral 
galaxies together with those of the giant HII regions (GHRs) observed by  
McCall, Rybski \& Shields (\cite{mcc}) in several spiral galaxies and
which 
compose the GHR
sequence. This sequence 
is interpreted as being a sequence in metallicity (O/H) and  
also in mean effective temperature or in mean ionization parameter (or
both)  
(McCall et al. \cite{mcc}, Dopita \& Evans \cite{do}). 
Note that radial abundance gradients, which are 
a common feature in spiral galaxies
(Zaritsky et al. \cite{z1}), complicate the interpretation 
of integrated spectra: 
the contribution of each 
annular region is weighted by the luminosities of the 
HII regions found there and by their spectral properties.

%\begin{figure*}
%\resizebox{\textwidth}{!}{\includegraphics{figure7.ps}}
\begin{figure}
\resizebox{\columnwidth}{!}{\includegraphics{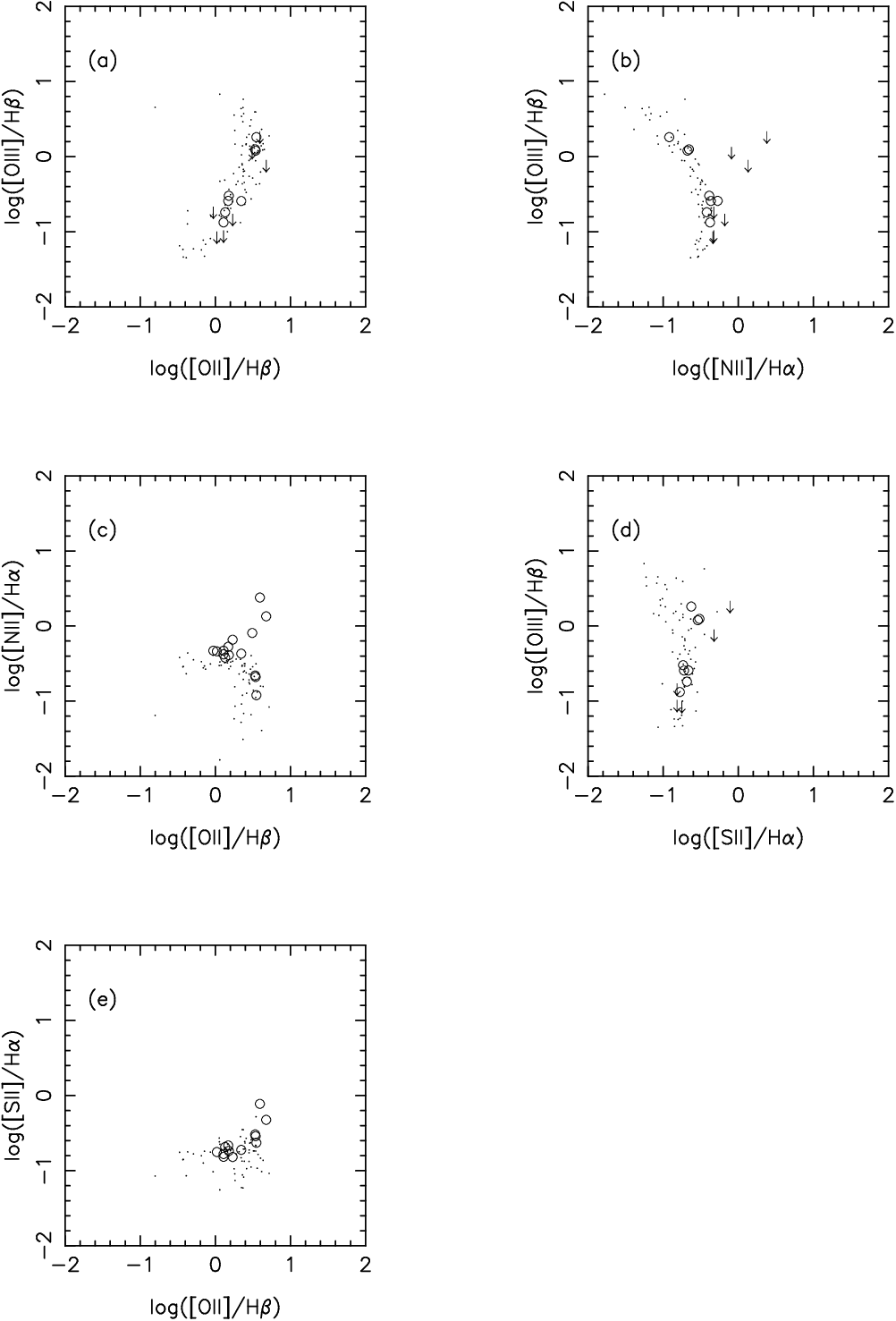}}
      \caption[]{
Classical emission line ratio diagnostic diagrams. The data for 
galaxies of our sample (circles) are plotted together with data
for the giant HII regions observed by  McCall et al. 
(\cite{mcc}) (dots).
Upper values (of [OIII]) are indicated by arrows.
The panels are:
{\bf a}) [OIII]/H$\beta$ versus [OII]/H$\beta$, {\bf b}) [OIII]/H$\beta$ 
versus [NII]/H$\alpha$, {\bf c}) [NII]/H$\alpha$ 
versus [OII]/H$\beta$, {\bf d}) [OIII]/H$\beta$ versus
[SII]/H$\alpha$, and {\bf e}) [SII]/H$\alpha$ versus
[OII]/H$\beta$.
                }
         \label{f7}
\end{figure}
%\end{figure*}

Panel 7-a shows [OIII]/H$\beta$ versus [OII]/H$\beta$. 
We see that the points corresponding to the integrated spectra of our
standard galaxies that have [OIII] large enough to be observed
(i.e. those with ST $>$0)  are well inside the GHR sequence. 

Panel 7-b displays  [OIII]/H$\beta$ versus [NII]/H$\alpha$.
Galaxies with measured [OIII] emission are disposed along the 
GHR sequence. The outliers are galaxies for which only upper values
 of the [OIII] emission are available. They are early type spirals, with 
ST $<$ 0, as can be deduced from Figs. 5c and 5d.
These galaxies stand out conspicuously in the [NII]/H$\alpha$ versus 
[OII]/H$\beta$, diagram (Panel 7c) as well.  As will be seen below, we 
interpret this behavior as being the combined effect of an 
overabundance of nitrogen in early-type spirals and an ionization 
source different from ordinary O stars.

Panel 7-d displays [OIII]/H$\beta$ versus [SII]/H$\alpha$.  In this
diagram, 
while the standard galaxies with 
measured $\log($ [OIII]/  H$\beta$) $<$ 1 are within the HII region 
sequence, 
those with $\log($ [OIII]/ H$\beta$) $>$1 
(i.e. all the galaxies with  ST $>$ 9) are slightly to the upper right. 
This may indicate an increasing contribution 
of a diffuse ionized medium with increasing galaxy spectral type. Note 
that Lehnert  \&  Heckman (\cite{leh}), plotting the 
spectra of K82a having  EW(H$\alpha$) $>$ 30 \AA~  in [OIII]/ H$\beta$
 versus [NII]/ H$\alpha$ and 
[OIII]/ H$\beta$ versus [SII]/ H$\alpha$ planes also concluded that 
diffuse ionized gas 
contributes to the integrated spectra of galaxies. \footnote{ 
The detection of a diffuse medium is actually not possible 
using a [OIII]/ H$\beta$
 vs [NII]/ H$\alpha$ diagram. Indeed, photoionization models (Stasi\'nska
\cite{st})
 shows that decreasing the ionization 
 parameter alone to very low values does not shift the points out of 
 the GHR sequence. On the other hand, this diagram is sensitive to the N/O
ratio
 which, unlike S/O, is expected to vary significantly from object to 
 object. Detailed spectroscopic observations of the diffuse intergalactic
 medium in spiral galaxies (Wang et al. \cite{wal}) however suggest that the 
ionization
 of this medium cannot be entirely attributed to radiation from O stars.}

Panel 7-e shows [OII]/H$\beta$ versus [SII]/H$\alpha$.
The same early-type spirals that show enhanced [NII] emission also show
enhanced [SII] emission compared with the GHRs sample of McCall et al.
\cite{mcc},
 although to a lesser extent. They are among the ones with the 
largest [OII]/H$\beta$ ratios. We will argue below that the 
bulk of the emission 
in these objects is not due to ordinary O stars.

%\begin{figure}
\begin{figure*}
%\resizebox{\columnwidth}{!}{\includegraphics{figure8.ps}}
\resizebox{15cm}{!}{\includegraphics{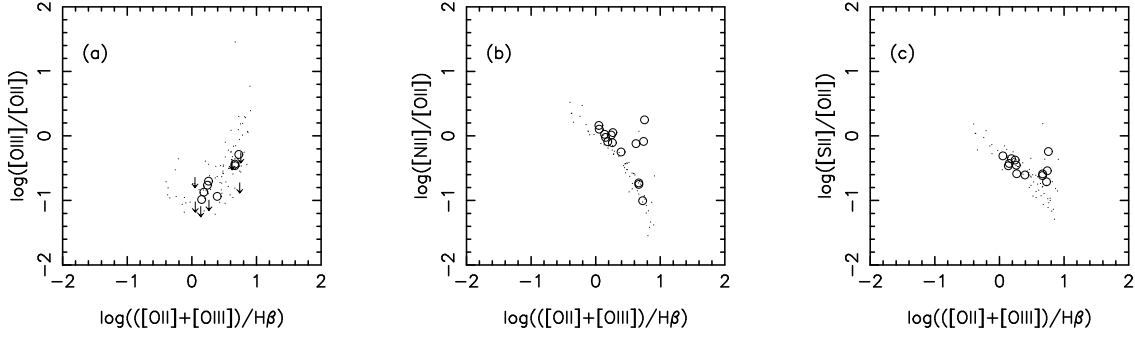}}
      \caption[]{
Forbidden line ratios as a function of
the O/H indicator ([OII]+[OIII])/H$\beta$. The data for 
galaxies of our sample (circles) are plotted together with data
for the giant HII regions observed by  McCall et al.
(\cite{mcc}) (dots).
Upper values (of [OIII]) are indicated by arrows.
The panels are:
{\bf a}) [OIII]/[OII] versus ([OII]+[OIII])/H$\beta$, {\bf b})
 [NII]/[OII] versus  ([OII]+[OIII])/H$\beta$, and {\bf c})
 [SII]/[OII] versus  ([OII]+[OIII])/H$\beta$.
                }
         \label{f8}
\end{figure*}
%\end{figure}

Fig. 8 shows some line ratios as a function of ([OII]+  [OIII])/ H$\beta$. 
Since the pioneering work of Pagel et al. (\cite{pa}), 
the ([OII]+ [OIII])/H$\beta$ ratio has been widely used to derive
the oxygen abundance. 
In principle, the relation between 
([OII]+[OIII])/ H$\beta$ and O/H is double sided (e.g. McGaugh \cite{mcg}, 
Stasi\'nska \cite{st}). 
In integrated spectra of galaxies,  regions within a few kiloparsecs from 
the galactic center have probably the greater weight, because most of the 
star formation is there (Rozas et al. \cite{roz}).
Since these regions are the most metal rich, one expects that 
the integrated spectra correspond to the regime where  
([OII]+ [OIII])/ H$\beta$ increases as O/H decreases (but this should be 
confirmed by simulations). 

Panel 8-a shows 
[OIII]/[OII] versus ([OII]+ [OIII])/ H$\beta$. [OIII]/ [OII] roughly
indicates the ionization state, which is linked to the 
ionization parameter $U$, thus to the gas density distribution in 
each HII region, and to the 
hardness of the ionizing  radiation field. 
In this diagram, the data points for the integrated galaxy spectra lie 
well within the region occupied by GHRs except, maybe, for the early
spectral type galaxies, with only upper values of [OIII]/ H$\beta$,
that tend to locate in the lower envelope of the GHR sequence.
The spread of ([OII]+ [OIII])/ H$\beta$ for the galaxies
is smaller than for individual GHRs. This 
is  reasonable,  since the integrated 
spectra are averages over GHRs of various metallicities, ionization 
parameters and mean affective temperatures. 

Panel 8-b displays [NII]/[OII] versus  ([OII]+ [OIII])/ H$\beta$.
We see that all galaxies in the sample fall inside the GHR sequence 
in fig. 8-b, except three  that 
have higher [NII]/ [OII] than GHRs of
 same ([OII]+ [OIII])/ H$\beta$. Note that the three exceptions also have
high [NII]/H$\alpha$ and are all the galaxies of our sample that have 
ST $<$ -3.

Panel 8-c displays  [SII]/[OII] versus ([OII]+[OIII])/H$\beta$. In this 
diagram, most of the standard galaxies fall inside the GHR sequence 
(although, as already noted by McCall et al. \cite{mcc}, the GHR 
sequence is not so well defined when using  [SII]/[OII]  instead of  
[NII]/[OII]), the galaxies with ST $<$ -3 tending to be above that
sequence.

\subsection{A tentative interpretation of emission line trends with 
galaxy spectral types}

While the trends between the emission line properties in the integrated 
spectra of normal spiral galaxies and spectral types are impressive,
their interpretation 
is not necessarily straightforward, since it is dependent of the physical
state
and spatial distribution of the gas and of the ionization mechanisms. 

It was shown by Zaritsky et al. 
(\cite{z1}) that the characteristic oxygen 
abundance in a spiral galaxy
derived from its individual giant HII regions increases towards  earlier
 morphological types.
The trend seen in  Fig. 5-f for ST $>$ 0 corresponds 
(at least qualitatively) to what is expected. 
Using the Zaritsky et al. (\cite{z1}) calibration of
([OII]+[OIII])/H$\beta$
into O/H,
it would translate into a decrease in the average O/H by
$\sim$0.5 dex for ST going from 
0 to 10. This is only an indicative value, because the
presence of abundance gradients makes it impossible to be more 
informative without numerical simulations.  
The case of the galaxies with ST $<$ 0, will be discussed later. 

Zaritsky et al. 
(\cite{z1}) have also found a strong correlation between 
the characteristic oxygen abundance
and galaxy luminosity (see also Vila-Costas \& Edmunds \cite{vi}; 
Roberts \& Haynes \cite{rob}).
Due to the strong correlation between galaxy luminosity and Hubble type in 
their sample, it is impossible to know which of these two quantities
is the primary cause of the correlation.
It is interesting to point out that, in the sample of galaxies discussed
here,  there is no such correlation between
O/H and luminosity (most of the galaxies in table 2 have M$_B$ 
in the narrow range between -19.9 and -20.7, see \cite{k92b})
and that the  correlation between luminosity and 
spectral type, if existent, is at most weak. This suggests
that O/H is rather linked to the stellar populations, as measured by
the spectral types, than to galaxy luminosity.

The increase of [OIII]/[OII] with  spectral type (Fig. 5-g) for 
ST $\simgreat$ 0 is probably due to an increase of  $<T_{eff}>$ 
(or an increase of $U$) when O/H decreases, similarly to what is 
invoked for individual GHRs (e.g., Garnett \& Shields \cite{ga}).

The increase of [NII]/[OII] with decreasing spectral type (Fig. 5-h) 
is probably due to an increase of N/O. Indeed, in
GHRs, the fact that [NII]/[OII] increases with decreasing 
([OII]+[OIII])/H$\beta$ is interpreted, using photoionization 
models, as being due to an increase of N/O with O/H
(Thurston, Edmunds \& Henry \cite{th}). 
It is likely  then, that in the spectral sequence of galaxies
there is an increase of the average N/O with decreasing spectral type, 
linked with the increase of O/H. 
This is in qualitative agreement with a secondary production of nitrogen.
{\it A priori}, one cannot exclude the fact that 
 the increase of [NII]/[OII] with 
decreasing spectral type could be simply attributed to the decrease of O/H
 (i.e., with N/O
staying constant), due to the thermal properties of GHRs.
Indeed, [NII]/[OII] = (N/O) ($\epsilon$(NII)/$\epsilon$(OII)), 
where $\epsilon$(NII) and $\epsilon$(OII) are the emissivities of the 
[NII] and [OII] lines, and $\epsilon$(NII)/$\epsilon$(OII)  
is a decreasing function of the electron temperature, thus 
an increasing function of O/H.
Note, however, that the three early-type galaxies that are out of the 
GHR sequence in Figs. 7-c, 7-e, and 8-b, must 
have particularly high N/O, because their large [NII]/[OII] cannot be 
due to a particularly low electron temperature, since [NII]/H$\alpha$ 
is large as well.

The ratio [SII]/[OII] does not present any significant trend with 
spectral type (Fig. 5-i). This is consistent with the assumption that 
the abundance ratio S/O is constant, since both S and O are primary
elements. 
There is a hint of a decrease in that line ratio with increasing 
spectral type but
this may be ascribed to the variations of the ionization parameter along
the sequence more than to variations in the relative abundances of S and
O.

The increase of [OII]/H$\beta$ as the galaxy spectral type decreases 
from 0 downwards (as well as that of [SII]/H$\alpha$ and [NII]/H$\alpha$, 
Figs. 5-b, d, e) is very interesting. Early spectral type  
galaxies appear as upper limits 
in the [OIII]/ H$\beta$ versus [OII]/H$\beta$
diagnostic diagram, since 
their [OIII] emission is weak. They stand out completely from the 
giant HII region sequence in diagnostic diagrams with [NII] or [SII]. 
The integrated spectra of these galaxies are similar 
to those of LINERs.

Interestingly, 11 of  the 15 galaxies of our sample have had their nuclear 
regions observed by Ho et al. (\cite{ho}). 
Their classification of these
nuclei into LINERs (L), Seyfert 2 (S2) and HII regions (H) is as
follows (see table 1): all the H type nuclei correspond to ST $>$ 0,
while NGC3623 and NGC3368 (ST $<$ -3) correspond to LINERs. 
NGC3147 (ST = 0.0) has a S2 nuclear
spectrum and NGC3627 (ST = 1.4) has a transition spectrum (T2/S2). 
The other galaxies in table 1 have not been observed by Ho et al. 
(\cite{ho}). So, the integrated spectra of the galaxies with earliest 
spectral types have characteristics 
of LINERs, and the nuclear regions of these galaxies too
(we have checked that the contribution of the LINER nuclei to the integrated
galaxy spectra is negligible). 
This seems to indicate that what gives rise to the LINER phenomenon, 
at least in the normal galaxies we are studying here, is not specifically 
related to the nucleus.

Given the evidence presented above that metallicity, as measured by O/H,
decreases along the  spectral sequence, 
it would be tempting to attribute the distinct behavior
of the galaxies having ST $<$ 0 in Figs. 5, 7, and 8 to 
over-abundances in the gas. But, as stated in Sect 4.1, over-abundances 
would produce low [OII]/H$\beta$, not large ones as observed.

Another possibility is that the distinctive features presented by 
early type galaxies are due to the aging of the stellar populations 
associated with GHRs.
Sodr\'e \& Cuevas (\cite{s97}), using the 
spectral synthesis code GISSEL (Bruzual \& Charlot \cite{bc}), have shown 
that the spectral sequence of galaxies is well reproduced if one 
assumes that the star formation rate has the form $\exp (-t/t_*)$, 
where $t_*$ is an increasing function of the galaxy spectral type.  
Such a model explains, at least qualitatively, why the equivalent 
width of H$\beta$ emission 
 increases with increasing galaxy spectral 
type, as found in Sect. 3.  
For galaxies with the earliest spectral types, most of the star formation 
has occurred long ago, and very few O stars are present.
In this case, the radiation field in the Lyman 
continuum may be dominated by post-AGB stars  (e.g. Bressan et al.  
\cite{bre}), because, in simple starburst models, these stars provide most
of
the ionizing photons for bursts older than $\sim 10^8$ yr.
This radiation field is much harder than that produced by
the massive  O stars that power classical HII regions. It is 
available to ionize the gas bound in the primitive HII regions 
or the diffuse material produced
by their eventual disruption.
Binette et al.  (\cite{bin}) have shown that the emission line properties
(EWs and line ratios) of early type galaxies  can be accounted for by 
photoionization by such a stellar population.  A similar model could 
well explain at the same time the high [OII]/H$\beta$, [SII]/H$\alpha$, 
and [NII]/H$\alpha$ found in the early spectral type galaxies of our 
sample.

The long lasting dispute about the 
ionization mechanism in LINERs - photoionization versus shocks - is not 
settled yet, precisely because under certain conditions both 
explanations can account for the observed emission line spectra.
Models of evolving stellar populations that estimate the 
energy release in supernovae and stellar winds (e.g., Leitherer \& Heckman 
\cite{lei}) will help in the solution of the problem.

We have seen that the effective extinction coefficient 
at H$\beta$, $C({\rm H}\beta)$, decreases steadily with increasing 
spectral type. 
The beautiful trend shown in Fig. 5-a is somewhat surprising, since
it is not commonly believed that early-type  spirals have higher
average extinction than galaxies of later types. Indeed, the 
related issue of the overall opacity in galaxies has been subject of 
significant debate, often with contradictory opinions (e.g., Valentjin
\cite{val}, Davies \& Burstein \cite{db}). 
Wang \& Heckman (\cite{wh}) have verified that the dust opacity, measured
from the ultraviolet to far-infrared luminosity ratio, increases with
increasing galaxy infrared luminosity. Taking into account that the
far-infrared luminosity is approximately constant for early morphological
type spirals and decreases slowly towards later morphological types
(e.g., Roberts \& Heynes \cite{rob}), the 
correlation noticed by Wang \& Heckman
might indicate that the dust opacity is indeed decreasing towards later
morphological types, in qualitative agreement with the results reported here.

Now, assuming for simplicity that dust properties are invariant with 
galaxy spectral type,  this trend 
suggests that the mean surface density of dust in galaxies decreases 
along the spectral sequence from early to late spectral types by a
factor of 4.  Since the mean surface density of
 neutral hydrogen in normal galaxies increases towards later types
(Roberts \& Heines
\cite{rob}), the dust-to-gas ratio must be decreasing as the spectral
 type increases.
This is consistent with the idea of
 dust being formed less efficiently in low metallicity environements, as 
shown observationally by Lisenfeld \& Ferrara (\cite{lf}).
Note that some of the scatter in Fig. 5-a may be produced
by the random inclination of the galaxies and that the far infrared properties
of normal galaxies actually suggest some variations of the dust properties 
 with galaxy type (Sauvage \& Thuan \cite{sau}).

\section{Summary}

We have analyzed the global emission line properties of normal spiral
galaxies
from their integrated spectra. Initially, we have selected 
from the Kennicutt's atlas (\cite{k92b}) 23
galaxies which do not present any sign of spectral or morphological
peculiarity. A PCA based only
on continuous and absorption features of the integrated spectra of these
galaxies enabled us to define the galaxy spectral types. 
We have then considered the subsample of 15 galaxies
whose integrated spectra in \cite{k92b} revealed emission lines. All 
these galaxies are spirals. We have studied the progression of the 
properties of these emission lines with galaxy spectral type.

Our major results are the following:
\begin{enumerate}
\item {The EWs of all the lines correlate well with EW(H$\alpha$). As a 
consequence, for practical purposes, the EW of any of 
the studied lines  ([OII]$\lambda$3727, H$\beta$, [OIII]$\lambda$5007,
 H$\alpha$, [NII]$\lambda \lambda$6548, 6583, and 
[SII]$\lambda \lambda$6717, 6731) can be used 
to estimate the average massive star formation rate, at least for galaxies
with spectral types $\simgreat$ 0. Galaxies with lower spectral type have
emission line spectra resembling low excitation LINERs and,
consequently, estimates of their global star formation rate based on the
their emission lines are  unreliable.}

\item {The EW of all lines correlate very well with ST, 
suggesting that they can be used
as indicators of galaxy spectral types. }

\item {Emission line ratios present a remarkable regularity when plotted
against galaxy spectral types. This stems from the fact that the strengths
of 
the emission lines are linked to the stellar populations, and the 
latter constitute the basis of the spectral classification of galaxies.
The plots are much noisier when using ordinary morphological types.}

\item {For galaxies with ST $\simgreat$ 0, the increase of 
(OII+ OIII)/ H$\beta$ with
increasing spectral type suggests a decreasing  average metallicity
along the spectral sequence.}

\item { The increase of [OIII]/[OII] with
 increasing spectral type may be due either to an increase of the mean
 effective temperature of the ionizing stars, 
$<T_{eff}>$, or to an increase of the characteristic ionization parameter
$U$.}

\item {The increase of [NII]/[OII] with decreasing spectral 
type is
probably due to an increase of the average N/O. 
This behavior is consistent with the fact that, in giant HII regions,  N/O 
increases with O/H. }

\item{The ratio [SII]/[OII] is essentially constant with spectral type
and is consistent with S/O being constant. This is expected because
S and O are both primary elements. }

\item {The integrated spectra of early-type spirals (ST $\simless$ 1)
are similar to those of LINERs. The long 
characteristic time for star formation 
in early-type spirals as derived by Sodr\'e \& Cuevas 
(\cite{s97}) provides a natural explanation in terms of ionization by
post-AGB stars. 
Interestingly, in several of these galaxies, the nuclei 
themselves have been characterized as 
dwarf LINERs in the work of Ho et al. (\cite{ho}).}

\item {The extinction by dust in HII regions seems to decrease with 
increasing spectral type. This may be a direct consequence of the 
decrease in mean metallicity along the spectral sequence.}
\end{enumerate}

While the interpretation of the trends found with spectral type 
is only tentative, the reality of these trends is beyond any doubt. 

This work confirms that the spectral classification is  useful in the
analysis of galaxy properties, being able to replace, with advantages,
the traditional morphological classification for many purposes.  
It produces an objective, quantitative and easy to implement
classification, 
that may be more satisfactory for the analysis of large data bases than 
the subjective and hard to obtain morphological types. 
Also, since spectra are closely linked to the physics of the stars and 
of the interstellar medium that
build up galaxies, spectral types are more adequate 
than morphological types for a quantitative approach
in studies of galaxy properties, as shown by this work.

\begin{acknowledgements}
LSJ benefited from  the support pro\-vi\-ded by  
FAPESP, CNPq and PRONEX/FINEP to his work, and warmly acknowledges the 
hospitality and support
of Observatoire de Meudon, where most of this work was realized. 
\end{acknowledgements}


\begin{thebibliography}{}
\bibitem[1981]{bal} Baldwin J. A., Phillips M. M., Terlevich R. J., 1981, 
PASP, 93, 5
\bibitem[1997]{bar} Barbaro G., Poggianti B. M., 1997, A\&A, 324, 490
\bibitem[1992]{bel} Belley J., Roy J. R., 1992, ApJS, 78, 61
\bibitem[1994]{bin} Binette L., Magris C. G., Stasi\'nska G., Bruzual A.
G., 
1994, A\&A, 292, 13
\bibitem[1998]{bro} Bromley B. C., Press W. H., Lin H., Kirshner R. P.
1998, 
preprint
\bibitem[1994]{bre} Bressan A., Chiosi C., Fagotto F., 1994, ApJS, 94, 63
\bibitem[1995]{bc} Bruzual G., Charlot S., 1995, Galaxy Isochrone
Synthesis
Spectral Evolution Library (GISSEL95), available from the authors
\bibitem[1995]{co} Connolly A. J., Szalay A. S., Bershady M. A., Kinney A.
L., 
Calzetti D. 1995, AJ, 110, 1071
\bibitem[1998]{cu} Cuevas H., Sodr\'e L., Capelato H. V.,
Quintana H., Proust D., 1998, in preparation
\bibitem[1995]{db} Davies J., Burstein D., 1995, The Opacity of Spiral
Disks, Dordrecht, Kluwer
\bibitem[1998]{lap} de Lapparent V., Galaz G., Arnouts S. 1998, preprint
\bibitem[1991]{vauc} de Vaucouleurs G., de Vaucouleurs A., Corwin H. 
G., Buta R. J., Paturel G., Fouqu\'e P., 1991, Third Reference Catalogue
of 
Bright Galaxies (RC3),
Springer-Verlag, Berlin Heidelberg New York
\bibitem[1986]{do} Dopita M. A., Evans, I. N., 1986, ApJ, 307, 431
\bibitem[1996]{fo} Folkes S. R., Lahav O., Maddox S. J., 1996, MNRAS, 283,
651
\bibitem[1998]{gala} Galaz G., de Lapparent V. 1998, A\&A, 332, 459
\bibitem[1989]{gall} Gallagher J. S., Bushouse H., Hunter D. A., 1989, 
AJ, 97, 700
\bibitem[1987]{ga} Garnett D. R., Shields G. A., 1987, ApJ, 317, 82
\bibitem[1980]{he} Heckman T., Balick B., Crane P., 1980, A\&A, 40, 295
\bibitem[1997]{ho} Ho L. C., Filippenko A.V., Sargent W. L. W., 1997, 
ApJS, 112, 391
\bibitem[1936]{hu} Hubble E., 1936, The Realm of Nebulae, Yale Univ.
Press, 
New Haven
\bibitem[1990]{is} Isobe T., Feigelson E. D., Akritas M. G., Babu G. J.,
1990,
ApJ, 364, 104
\bibitem[1983]{keel} Keel  W. C., 1983, ApJS, 52, 229
\bibitem[1983]{k83} Kennicutt R. C., 1983, ApJ, 272, 54
\bibitem[K92a]{k92a} Kennicutt R. C. 1992a, ApJ, 388, 310
\bibitem[K92b]{k92b} Kennicutt R. C. 1992b, ApJS, 79, 255
\bibitem[1994]{k94} Kennicutt R. C., Tamblyn P., Congdon C. W., 1994, 
ApJ, 435, 22
\bibitem[1996]{k96} Kennicutt R. C., Garnett D. R., 1996, ApJ, 456, 504
\bibitem[1995]{lei} Leitherer C., Heckman T. M., 1995, ApJS, 96, 9 
\bibitem[1994]{leh} Lehnert M. D., Heckman T. M., 1994, ApJ, 426, L27
\bibitem[1998]{lf} Lisenfeld U., Ferrara A., 1998, ApJ 496, 145
\bibitem[1985]{mcc} McCall M. L., Rybski P. M., Shields G. A., 1985, 
ApJS, 57, 1
\bibitem[1991]{mcg} McGaugh S. S., 1991, ApJ, 380, 140
\bibitem[1993]{oe} Oey M. S., Kennicutt R. C., 1993, ApJ, 411, 137
\bibitem[1989]{os} Osterbrock D., 1989, Astrophysics of Gaseous Nebulae 
and Active Galactic Nuclei, University Science Books, Mill Valley, Ca.
\bibitem[1979]{pa} Pagel B. E. J., Edmunds M. G., Blackwell D. A., Chun
M. S., Smith G., MNRAS, 189, 95
\bibitem[1994]{rob} Roberts M. S., Haynes M. P., 1994, ARAA, 32, 115
\bibitem[1996]{roz} Rozas M., Knapen J. H., Beckman J. E., 1996, A\&A,
312, 275
\bibitem[1997]{roy} Roy J. R., Walsh J. R., 1997, MNRAS, 288, 715
\bibitem[1961]{san} Sandage A., 1961, The Hubble Atlas of Galaxies, 
Carnegie Institute of Washington, Washington, DC
\bibitem[1994]{sau} Sauvage M., Thuan T. X., 1994, ApJ, 429, 153
%\bibitem[1995]{shi} Shields J.C., Kennicutt R. C., 1995, ApJ, 454, 807
\bibitem[1994]{s94} Sodr\'e L., Cuevas H. 1994, Vistas in Astron., 38, 287
\bibitem[1997]{s97} Sodr\'e L., Cuevas H. 1997, MNRAS, 287, 137
\bibitem[1998]{st} Stasi\'nska G., 1998, in ``Dwarf Galaxies and
Cosmology", 
eds. Cayatte \& Thuan, Editions Frontieres, in press
\bibitem[1996]{th} Thurston T. R., Edmunds M. G., Henry R. B. C., 1996,
MNRAS, 283, 990
\bibitem[1990]{val} Valentijn E. A., 1990, Nat., 346, 153
\bibitem[1987]{ve} Veilleux S., Osterbrock D., 1987, ApJS, 63, 285
\bibitem[1992]{vi} Vila-Costas M. B., Edmunds M. G., 1992, MNRAS, 259, 121
\bibitem[1996]{wh} Wang B., Heckman T.M., 1996, ApJ, 457, 645
\bibitem[1997]{wal} Wang J., Heckman T.M., Lehnert M.D., 1997, ApJ, 491, 114
\bibitem[1994]{z1} Zaritsky D., Kennicutt R. C., Huchra J. P., 1994, 
ApJ, 420, 87
\bibitem[1995]{z2} Zaritsky D., Zabludoff A. I., Willick J. A. 1995, 
AJ, 110, 1602

\end{thebibliography}
\end{document}